\documentclass[aps,prl,superscriptaddress,reprint,nofootinbib]{revtex4-1}

\usepackage{graphicx,multirow}
\usepackage{bm}
\usepackage{amsmath}
\usepackage{amssymb}
\usepackage{amscd}
\usepackage{latexsym}
\usepackage{slashed}
\usepackage{color}
\usepackage{graphicx}
\usepackage{ulem}
\usepackage{color}
\usepackage[caption=false]{subfig}
\usepackage{endnotes}

\def\ltap{\raisebox{-.6ex}{\rlap{$\,\sim\,$}} \raisebox{.4ex}{$\,<\,$}} 
\def\gtap{\raisebox{-.6ex}{\rlap{$\,\sim\,$}} \raisebox{.4ex}{$\,>\,$}}

\usepackage{lineno}

\begin{document}


\title{
{Longitudinal $\mathbf{Z}$-Boson Polarization and the Higgs Boson Production Cross Section\\ at the Large Hadron Collider}
}

\author{S.~Amoroso}%
\email{simone.amoroso@desy.de}
\affiliation{Deutsches Elektronen-Synchrotron DESY, D 22607 Hamburg}%
\author{J.~Fiaschi}%
\email{fiaschi@uni-muenster.de}
\affiliation{Department of Mathematical Sciences, University of Liverpool, Liverpool L69 3BX}%
\affiliation{Institut f\"ur Theoretische Physik, Universit\"at M\"unster, D 48149 M\"unster}%
\author{F.~Giuli}%
\email{francesco.giuli@roma2.infn.it}
\affiliation{CERN, CH-1211 Geneva 23, Switzerland}
\affiliation{University of Rome Tor Vergata and INFN, Sezione di Roma 2, 00133 Roma}
\author{A.~Glazov}%
\email{alexander.glazov@desy.de}
\affiliation{Deutsches Elektronen-Synchrotron DESY, D 22607 Hamburg}%
\author{F.~Hautmann}%
\email{hautmann@thphys.ox.ac.uk}
\affiliation{Elementaire Deeltjes Fysica, Universiteit Antwerpen, B 2020 Antwerpen}
\affiliation{Theoretical Physics Department, University of Oxford, Oxford OX1 3PU}%
\author{O.~Zenaiev}
\email{oleksandr.zenaiev@cern.ch}
\affiliation{CERN, CH-1211 Geneva 23, Switzerland}
\begin{abstract}
\noindent
Charged lepton pairs are produced copiously in high-energy hadron collisions via electroweak gauge boson exchange, and 
are one of the most precisely measured final states in proton-proton collisions at the  Large Hadron Collider (LHC). 
We propose that measurements of lepton angular distributions can be used to improve the accuracy of theoretical 
predictions for Higgs boson production cross sections at the LHC.  
To this end, we exploit the sensitivity of the lepton angular coefficient associated with the longitudinal 
$Z$-boson polarization to the parton density function (PDF) for gluons resolved from the incoming protons, in order to constrain the Higgs boson cross section from gluon fusion processes.  
By a detailed numerical analysis using the open-source platform {\tt{xFitter}}, we find that high-statistics determinations of the longitudinally polarized angular coefficient at the LHC Run III and high-luminosity HL-LHC improve the PDF systematic uncertainties of the Higgs boson cross section predictions  by 50  $ \% $ over a broad range of Higgs boson rapidities.    
\end{abstract}

\vspace*{-1.5cm} 
\hspace*{1.5cm} DESY 20-216, MS-TP-20-44, LTH 1248 

\maketitle

{\it Introduction.} Precision studies in the Higgs sector of the Standard Model (SM) are central to 
current~\cite{deFlorian:2016spz} and forthcoming~\cite{Cepeda:2019klc} physics 
programs at the Large Hadron Collider (LHC), and provide a portal to searches 
for  beyond-Standard-Model (BSM) physics. The dominant mechanism for the 
production of Higgs bosons in proton-proton collisions at the LHC is given by the 
fusion of two gluons resolved from the incoming protons. With the very high 
accuracy reached in perturbative Quantum Chromodynamics (QCD) calculations of 
gluon-initiated production cross sections, currently of next-to-next-to-next-to-leading 
order (N$^3$LO)~\cite{Chen:2021isd,Dulat:2018bfe,Anastasiou:2016cez} in the QCD coupling $\alpha_s$,  
the theoretical systematic uncertainties affecting the predictions for gluon fusion 
processes receives  essential contributions from the non-perturbative gluon parton density function (PDF), as well as the 
sea-quark densities  coupled to gluons through   initial-state QCD evolution.  See e.g.~\cite{Cepeda:2019klc}, where the 
PDF  contribution is estimated to be about 30  $ \% $ of the total uncertainty, including $\alpha_s$ and scale variations.

The primary source of knowledge of the gluon PDF is given at present, in global fits to hadron collider 
data~\cite{Hou:2019efy,Ball:2017nwa,Alekhin:2017kpj,Harland-Lang:2014zoa,Ball:2014uwa,Abramowicz:2015mha,Bailey:2020ooq}, by deep inelastic scattering (DIS) experimental measurements at high energy.  
Future DIS experiments~\cite{Agostini:2020fmq,Aidala:2020eah} are proposed to extend the range and accuracy of our current knowledge of the gluon PDF. It is hoped that substantial progress can also come from measurements at the LHC itself, particularly in the forthcoming high-luminosity phase HL-LHC~\cite{Azzi:2019yne}.  
Gluon PDF determinations are considered  from open~\cite{Zenaiev:2019ktw,Cacciari:2015fta} and bound-state~\cite{Flett:2019pux} charm and bottom quark production, light-quark jets~\cite{AbdulKhalek:2020jut} and top quark production~\cite{Czakon:2016olj}.

In this work we take color-singlet hadro-production (unlike  the above cases, in which the Born-approximation final state contains colored particles) and, similarly to the case of DIS, investigate the sensitivity to the gluon PDF via ${\cal O} ( \alpha_s)$ contributions, guided by criteria of perturbative stability and experimental precision. 
 
We consider Drell-Yan (DY) charged lepton-pair production~\cite{Drell:1970wh} via electroweak vector boson exchange.  
Let us map the DY cross section in the boson invariant mass $M$, rapidity $Y$ and transverse momentum $p_T$, and in the lepton polar and azimuthal angles $\theta$ and $\phi$, defined in the Collins-Soper reference frame~\cite{Collins:1977iv}  (see e.g.~the DY cross section parameterization in~\cite{Aad:2016izn}).    
The DY cross section summed over the electroweak boson polarizations has the angular distribution $1 + \cos^2 \theta$ and is sensitive to the gluon PDF for finite $p_T$. However, in the $p_T$ region where the cross section is the largest, it is affected by large perturbative corrections to all orders in $\alpha_s$ (see e.g.~\cite{Angeles-Martinez:2015sea}, and references therein). Let us turn to contributions of the single electroweak-boson polarizations. The diagonal 
elements of the polarization density matrix~\cite{Collins:1977iv,Lam:1978pu,Chaichian:1981va,Hagiwara:1984hi} in the helicity basis yield  (besides  the unpolarized cross section, proportional to the trace of the density matrix) the forward-backward asymmetry and the longitudinally polarized cross 
section, associated respectively with angular distributions $ \cos \theta$ and $(3 \cos^2 \theta -1) /2$. The former is parity-violating and dominated by flavor non-singlet PDFs~\cite{Abdolmaleki:2019qmq,Abdolmaleki:2019ubu,Accomando:2018nig,Accomando:2017scx,Fiaschi:2019skn,Accomando:2019ahs,Fiaschi:2018buk}. 
The latter is parity-conserving and  sensitive to flavor singlet PDFs. 
Off-diagonal density matrix elements can be accessed by measuring, besides $\theta$, the lepton's azimuthal angle $\phi$, and yield six additional contributions besides the previous three, leading to nine linearly independent polarized cross sections, which correspond to the first nine terms in the expansion over  spherical harmonics. 

In order to constrain the Higgs boson production cross section from gluon fusion, we will focus on the 
ratio of the longitudinal  electroweak boson cross section to the unpolarized cross section, defining the angular coefficient  
\begin{equation} 
\label{A0def} 
A_0 ( s , M, Y, p_T) = { {2 d \sigma^{(L)} / dM dY dp_T } \over   { d \sigma / dM dY dp_T} } . 
\end{equation}   
The coefficient $A_0$ is  perturbatively stable,  as illustrated by the smallness of its 
 next-to-leading-order (NLO)~\cite{Mirkes:1992hu,Mirkes:1994eb,Mirkes:1994dp,Lambertsen:2016wg,Peng:2015spa,Chang:2017kuv} and  
next-to-next-to-leading-order (NNLO)~\cite{Gauld:2017tww} radiative corrections for finite $p_T$, and 
precisely measured at the LHC~\cite{Khachatryan:2015paa,Aad:2016izn}, following earlier measurements at the 
Tevatron~\cite{Aaltonen:2011nr} 
and fixed-target experiments~\cite{Falciano:1986wk,Guanziroli:1987rp,Conway:1989fs,Zhu:2006gx,Zhu:2008sj}. 
We will comment later on the extension of the analysis to other polarized contributions besides the longitudinal cross section.

We now proceed as follows. First, we discuss the general properties of the angular coefficient (\ref{A0def}) illustrating the physics potential of precision measurements of DY angular distributions at the LHC Run III and HL-LHC. Next, we focus on its application to the profiling of the gluon distribution using the open-source fit platform {\tt{xFitter}}~\cite{Alekhin:2014irh}, and compute the resulting Higgs boson cross section and PDF uncertainty at $\sqrt{s} = 13$~TeV.

\begin{figure}[t!]
\begin{center}
\includegraphics[width=0.30\textwidth,angle=270]{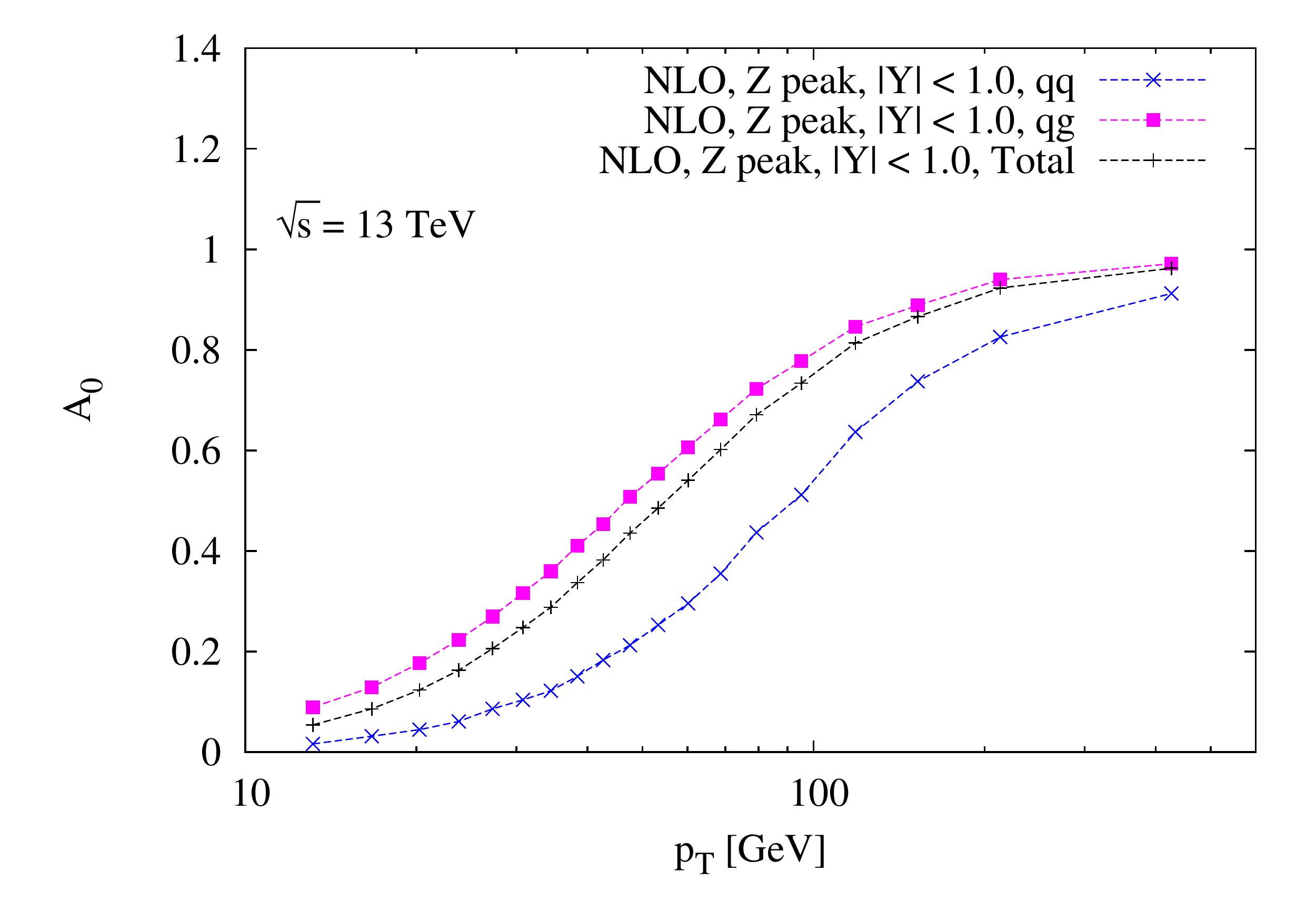}
\includegraphics[width=0.30\textwidth,angle=270]{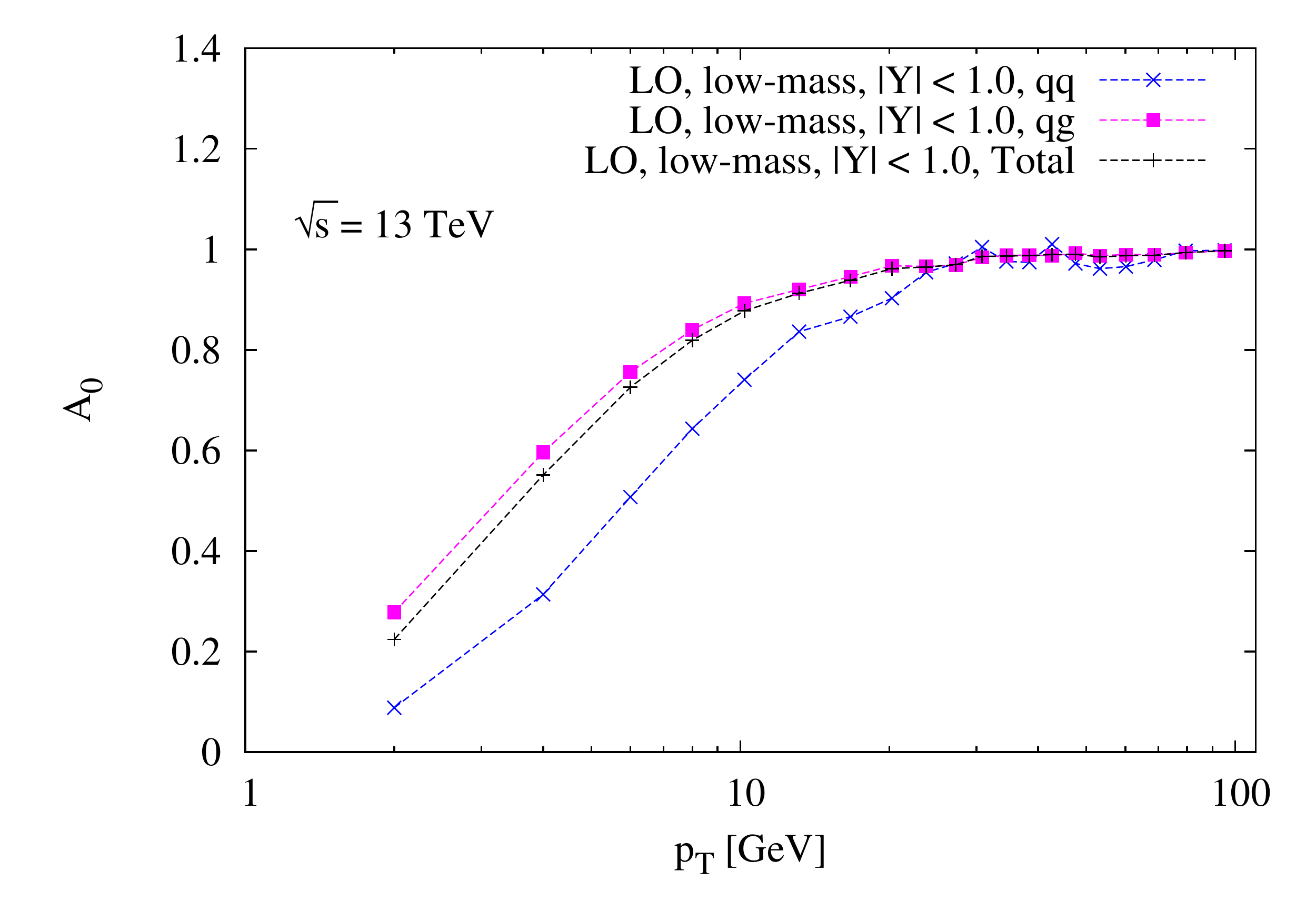}
\includegraphics[width=0.30\textwidth,angle=270]{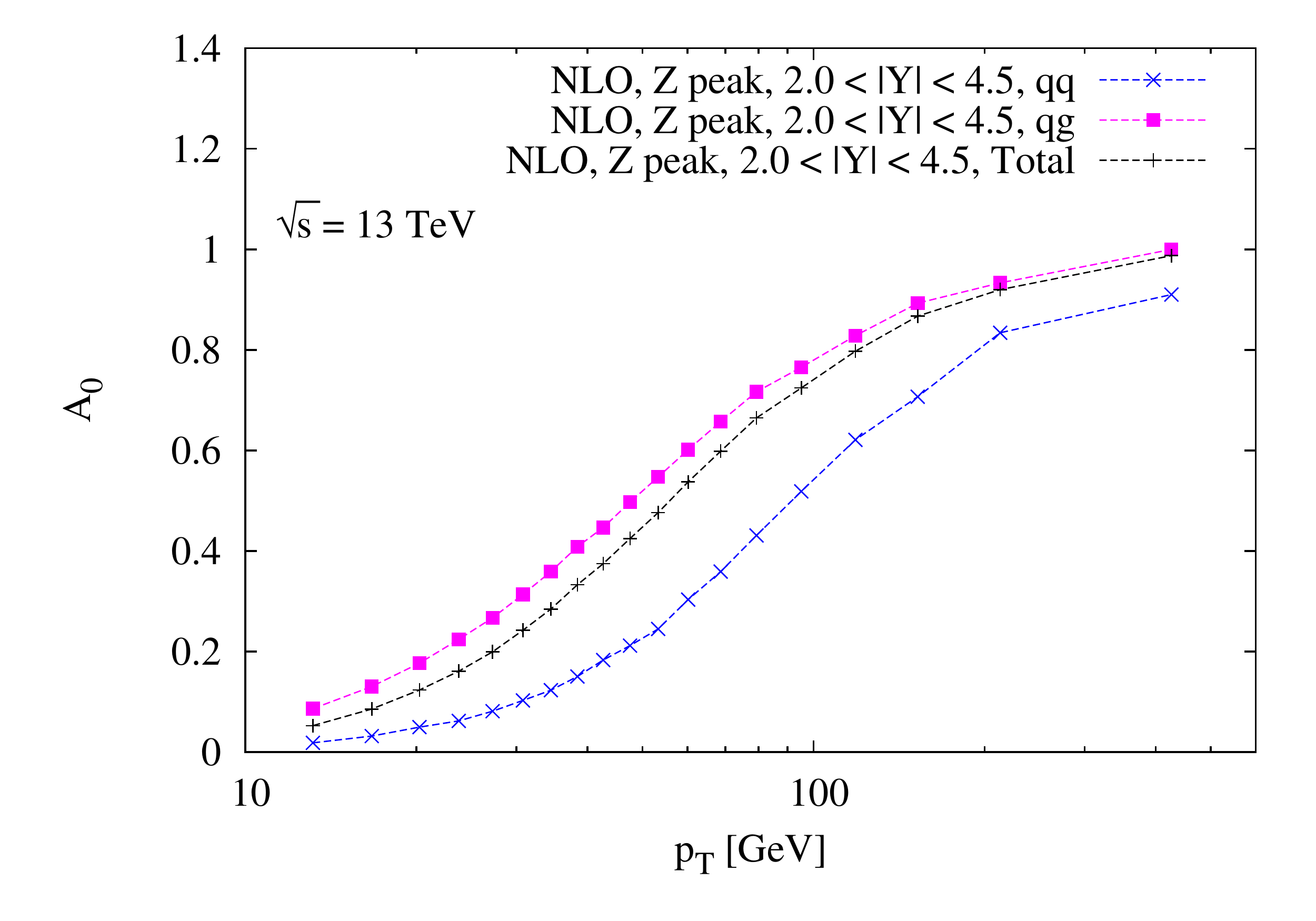}
\caption{The angular coefficient $A_{0}$ and its  $q\bar{q}$, $qg$ contributions for $\sqrt{s} = 13$ TeV as functions 
of the boson $p_T$  based on CT18nnlo PDF set. The results are plotted in different regions of the boson invariant mass $M$ and rapidity $Y$: the $Z$-boson peak region, 80 GeV $< M <$ 100 GeV, for $|Y| <$ 1.0 (NLO, top); low-mass region between the $ J/\Psi$ and  $\Upsilon$ resonances, 4 GeV $< M <$ 8 GeV, for $|Y| <$ 1.0 (LO, center); $Z$-boson peak region, 80 GeV $< M <$ 100 GeV for LHCb kinematics (NLO, bottom).  
 }
\label{fig:A0_13TeV_contributions}
\end{center}
\end{figure}

{\it Longitudinally polarized angular coefficient.} 
The longitudinally polarized coefficient $A_0$ in Eq.~(\ref{A0def}) vanishes in the parton model and 
receives leading-order (LO) perturbative QCD contributions at ${\cal O} (\alpha_s)$.  
We evaluate $A_0$ at LO and NLO (i.e., through ${\cal O} (\alpha_s^2)$) using the {\tt{MadGraph5\_aMC@NLO}}~\cite{Alwall:2014hca} program 
for $Z$ + 1 parton $pp$ production.   
In Fig.~\ref{fig:A0_13TeV_contributions}\footnote{While no cut is applied on the parton $p_T$, a cut on the $Z$-boson $p_T$ of 11.4 GeV is used 
for calculations in the $Z$-boson mass region. This is lowered to 1 GeV for the low-mass region.} 
we show results for $A_0$ at the energy $\sqrt{s} = 13$ TeV versus the boson $p_T$ 
for three distinct kinematic regions: two of them at central rapidity $|Y|<1$ with invariant mass $M$ close to 
the $Z$ boson mass $M_Z$ as well as between the $J/\psi$ and $\Upsilon$ resonances, corresponding to ATLAS and CMS kinematics; and one with $2 < |Y| < 4.5$ and $M$ close to $M_Z$, corresponding to LHCb kinematics.
We also show separately the contributions to $A_0$ from the  initial-state partonic channels $q\bar{q}$ and  $qg$  ($qq$ and $gg$ channels are present at NLO, but they are suppressed by relative order $\alpha_s$).  

The coefficient $A_0$ as well as its $q\bar{q}$ and  $qg$ components rise monotonically from 0 at $p_T = 0$ to 1 for $ p_T \gg M$.
The gluonic channel $qg$ dominates over the fermionic channel $q\bar{q}$, for the low mass region in particular. The weight of the $q\bar{q}$ contribution increases for the $Z$ pole region and  reaches its largest value for the LHCb phase space.
This can be understood since the range in longitudinal momentum fraction $x$ probed for the PDFs is changing from low to high $x$. 
The location of $A_0$ measured in experimental data with respect to the predicted $q\bar{q}$ and $qg$ contributions can constrain the $\bar{q}/g$ ratio.

The main sensitivity to the gluon distribution arises from the $A_0$ region with the largest slope in $p_T$, i.e., around the turn-over point $\partial^2 A_0 / \partial p_T^2=0$. 
Figure~\ref{fig:A0_13TeV_contributions} illustrates that the position of this turn-over point varies strongly with a power-like dependence  on the lepton-pair invariant mass, so that the mass provides a powerful handle on the $p_T$ scales probed in the initial state distribution. 
Near the $Z$-boson peak the turn-over occurs at $p_T$ of the order of several ten to 100~GeV, while at low masses, between the $ J/\Psi$ and $\Upsilon$ meson resonances, it is at $p_T$ of the order of a few~GeV.
This behavior is to be contrasted with the case of $d\sigma / dp_T$,  which peaks at low  $p_T$, with the position of the peak depending only very mildly on the invariant 
mass~\cite{Collins:1977iv,Angeles-Martinez:2015sea}. 
Thus, we will use the longitudinally polarized angular distribution near the electroweak boson mass scale in order to constrain the gluon PDF in the region relevant~\cite{Cipriano:2013ooa} for Higgs boson production (as will be described  next).   
On the other hand, the above observations on the meson resonance region suggest  that the angular distribution in this region  
can provide sensitivity to the $p_T$ dependent PDFs~\cite{Angeles-Martinez:2015sea,Motyka:2016lta,Boer:2017hqr,Hautmann:2020cyp} (which we leave to future investigations).

New features therefore arise in the extraction of non-perturbative QCD contributions  
owing to  the vector boson polarization.  
%
In what follows, we  carry out a detailed analysis for collinear distributions.

\begin{figure}[t!]
\begin{center}
\includegraphics[width=0.23\textwidth]{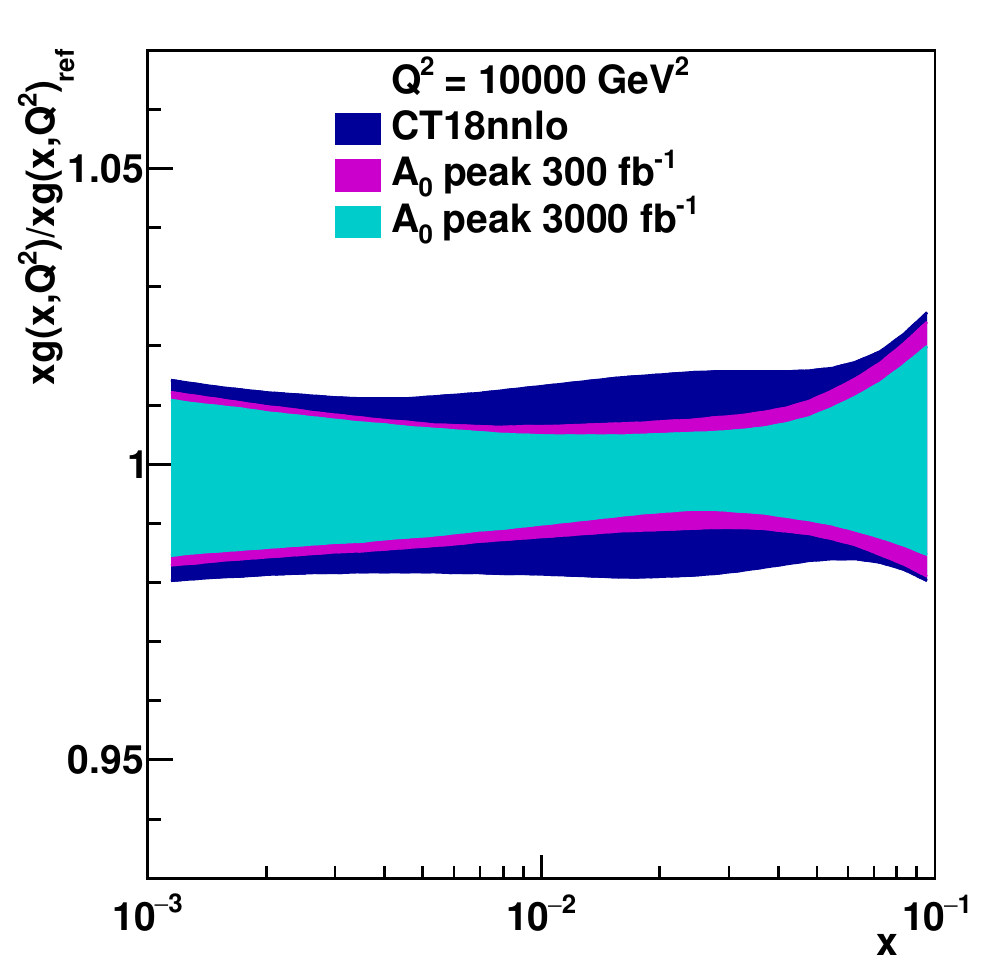}
\includegraphics[width=0.23\textwidth]{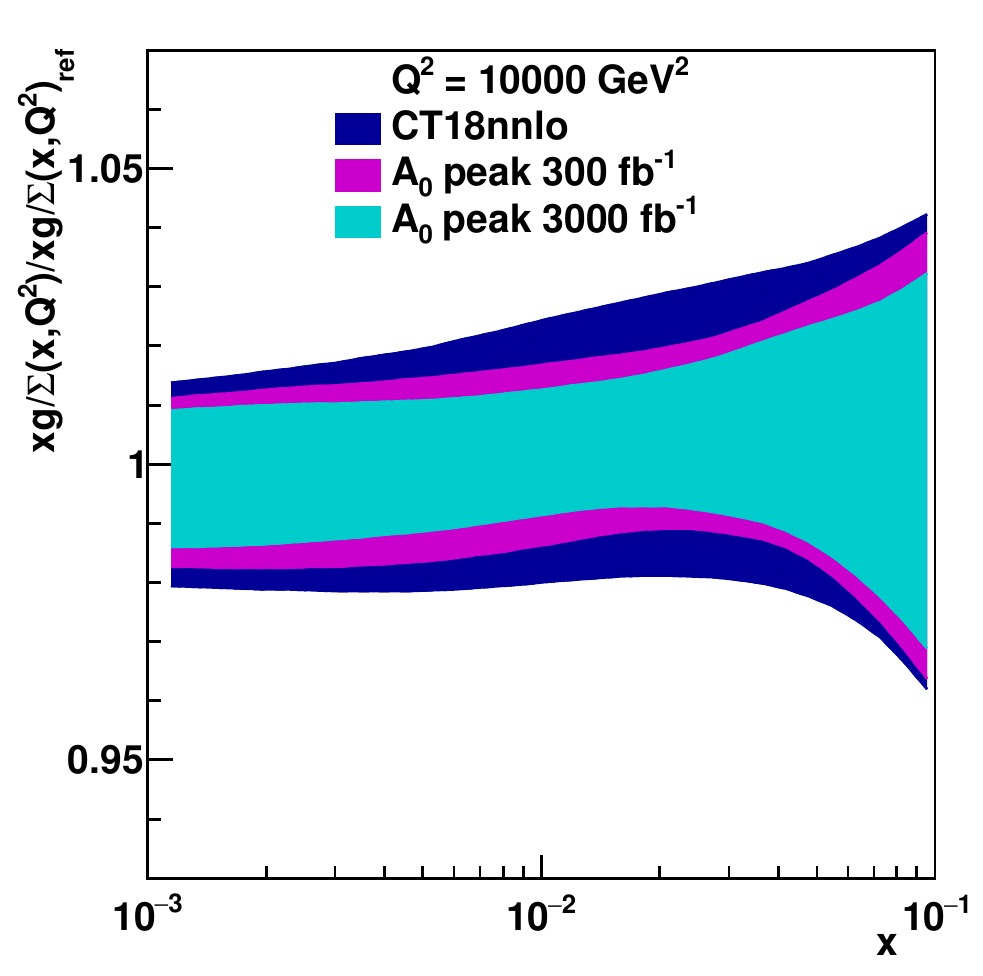}
\includegraphics[width=0.23\textwidth]{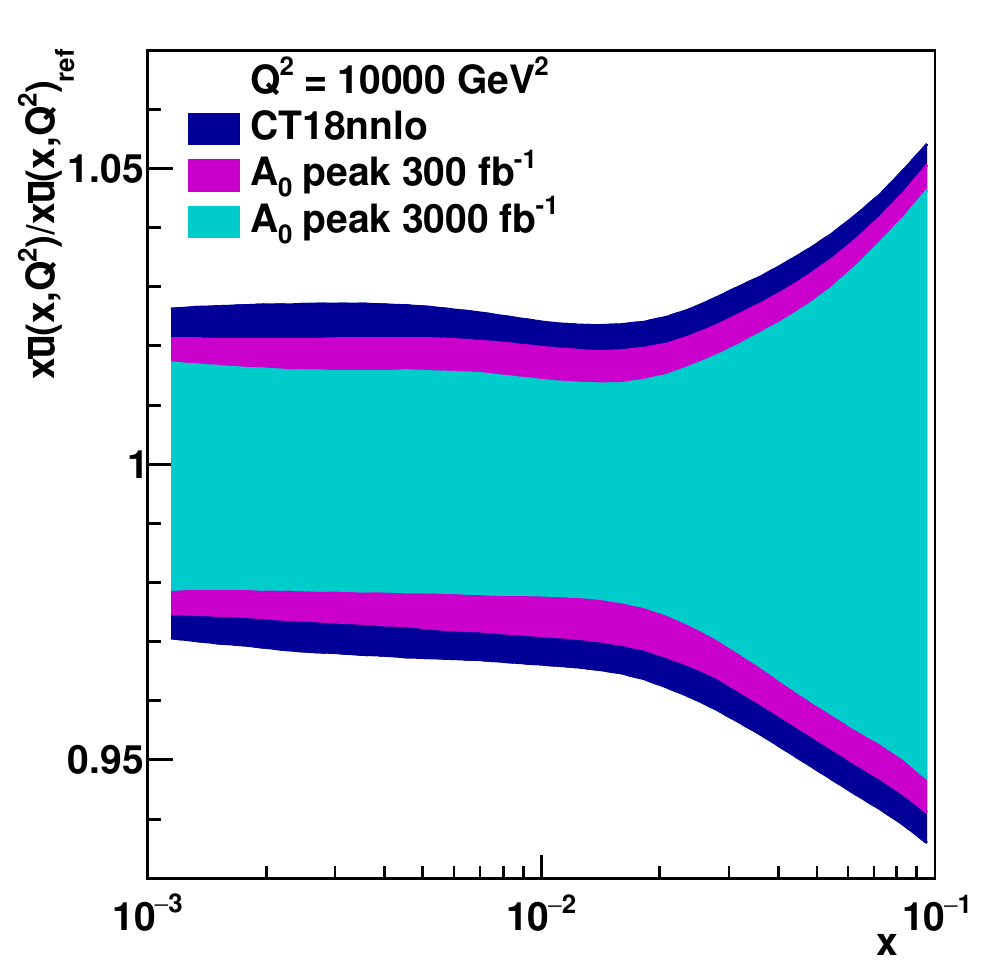}
\includegraphics[width=0.23\textwidth]{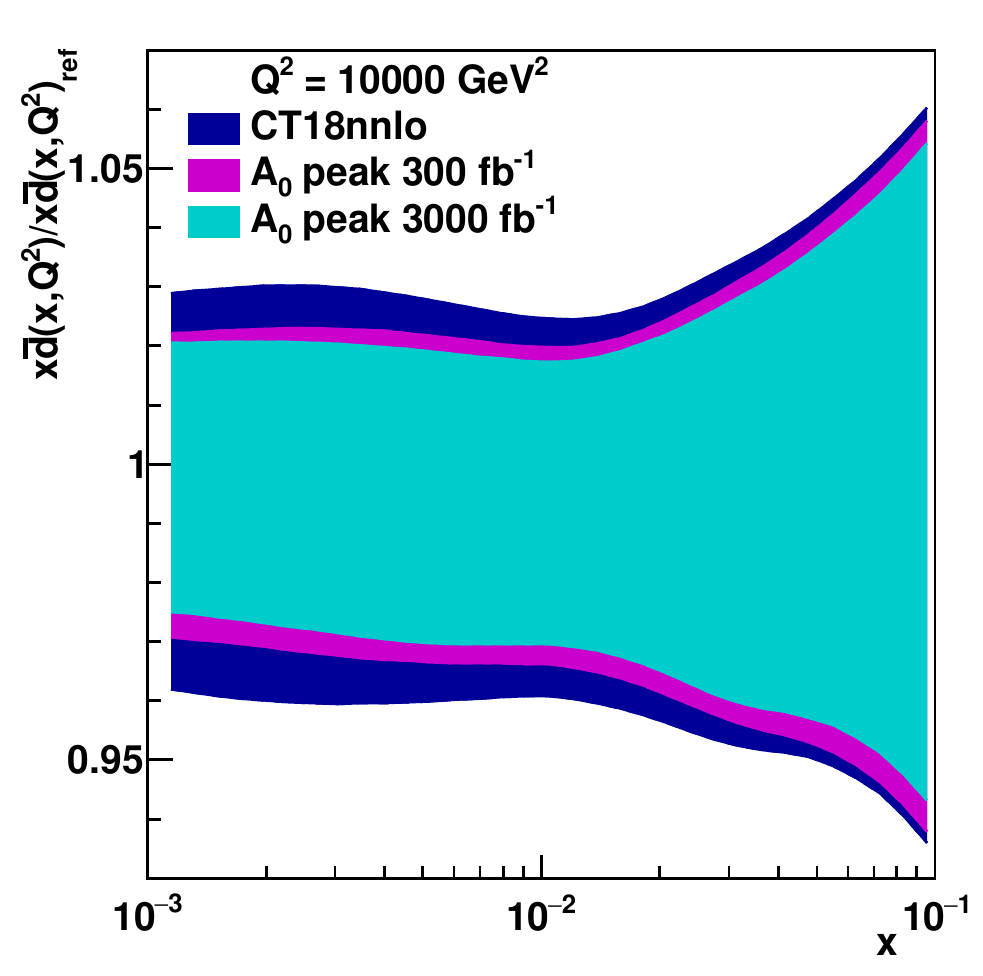}
\caption{Original CT18nnlo~\cite{Hou:2019efy} (red) and profiled distributions using $A_0$ pseudodata corresponding to integrated luminosities of 300~fb$^{-1}$ (blue) and 3~ab$^{-1}$ (green) for 80~GeV $< M <$ 100~GeV and $|Y|<3.5$. Results for gluon ($xg$), gluon/Sea ($xg/\Sigma$), $u$-type ($x\bar{u}$) and $d$-type ($x\bar{d}$) sea-quark densities are shown for $Q^{2}$ = 10$^{4}$~GeV$^{2}$. Bands represent PDF uncertainties, shown at the 68\% CL.}
\label{fig:CT18_13TeV_Second}
\end{center}
\end{figure}

{\it Gluon profiling and Higgs cross section.} 
To analyze the  impact of high-precision $A_0$ measurements on the 
Higgs boson production cross section, we implement 
the NLO  {\tt{MadGraph5\_aMC@NLO}} $A_0$ calculation into the fit platform {\tt{xFitter}}~\cite{Alekhin:2014irh}.  
First, we validate our implementation by performing NLO fits to    
the $\sqrt{s} = 8$ TeV ATLAS measurements~\cite{Aad:2016izn} of $A_{0}$, and verifying that 
 good $\chi^2$ values are obtained for all the PDF sets considered, namely CT18nnlo~\cite{Hou:2019efy}, 
NNPDF3.1nnlo~\cite{Ball:2017nwa}, ABMP16nnlo~\cite{Alekhin:2017kpj},  HERAPDF2.0nnlo~\cite{Abramowicz:2015mha} and MSHT20nnlo~\cite{Bailey:2020ooq}. 
Next, we generate $A_0$ pseudodata for Z $p_T >$ 11.4 GeV at  $\sqrt{s} = 13$~TeV for two projected luminosity scenarios of 
300 fb$^{-1}$ (the designed integrated luminosity at the end of the LHC Run III) and 3 ab$^{-1}$ (the designed integrated luminosity at the end of the HL-LHC stage~\cite{Azzi:2019yne}), and apply  
the profiling technique~\cite{Paukkunen:2014zia,Camarda:2015zba} to evaluate the PDF uncertainties. 
To do this, we extrapolate the statistical uncertainties  for the two projected integrated luminosities, and estimate the systematic uncertainties assuming a 0.1\% systematic uncertainties in the lepton momentum scale.
\footnote{Note that PDF uncertainties are large in the ATLAS 8 TeV $A_0$ measurements~\cite{Aad:2016izn} extrapolated in rapidity $Y$, but they are small for the measurements~\cite{Aad:2016izn} in bins of $Y$.}
We perform the analysis in the  mass region  80~GeV $< M <$ 100~GeV around the $Z$-boson peak 
and  rapidity region $|Y| <$ 3.5.
The results are reported in Fig.~\ref{fig:CT18_13TeV_Second}.

We find that, in accord with the earlier discussion, the largest reduction of uncertainties from the high-luminosity $A_0$ profiling occurs for the gluon density 
(top two panels in Fig.~\ref{fig:CT18_13TeV_Second}), 
 and for the $u$ and $d$ sea-quark densities coupled to gluons through QCD evolution 
 (bottom two panels in Fig.~\ref{fig:CT18_13TeV_Second}). All panels in 
 Fig.~\ref{fig:CT18_13TeV_Second} show the range  $10^{-3} \ltap x \ltap 10^{-1}$ where the reduction is most pronounced. 
We  find that the largest sensitivity comes from transverse momenta in the mid range $p_T \sim 50$~GeV, and the sensitivity dies out for $p_T \gtap 100$~GeV. 
 The  current-to-300 fb$^{-1}$ gain dominates  the  300 fb$^{-1}$ to 3 ab$^{-1}$ gain, similarly to other earlier profiling examples (see detailed discussions  
in~\cite{Abdolmaleki:2019qmq} for valence quarks and in~\cite{Rojo:2015acz} for gluons).  

   We have also verified the perturbative stability of our results, by repeating the profiling with a variation of the perturbative factorization and renormalization scales at NLO
in the pseudodata. The central value for the resulting gluon distribution function stays within one standard deviation band of the profiled PDF uncertainty. Given that existing NNLO
predictions have significantly reduced scale uncertainty~\cite{Gauld:2017tww},  we expect that the effect of higher-order corrections will have only a small impact on the profiled PDFs, and this 
uncertainty is neglected in the following. 

\begin{figure}[t!]
\begin{center}
\includegraphics[width=0.43\textwidth]{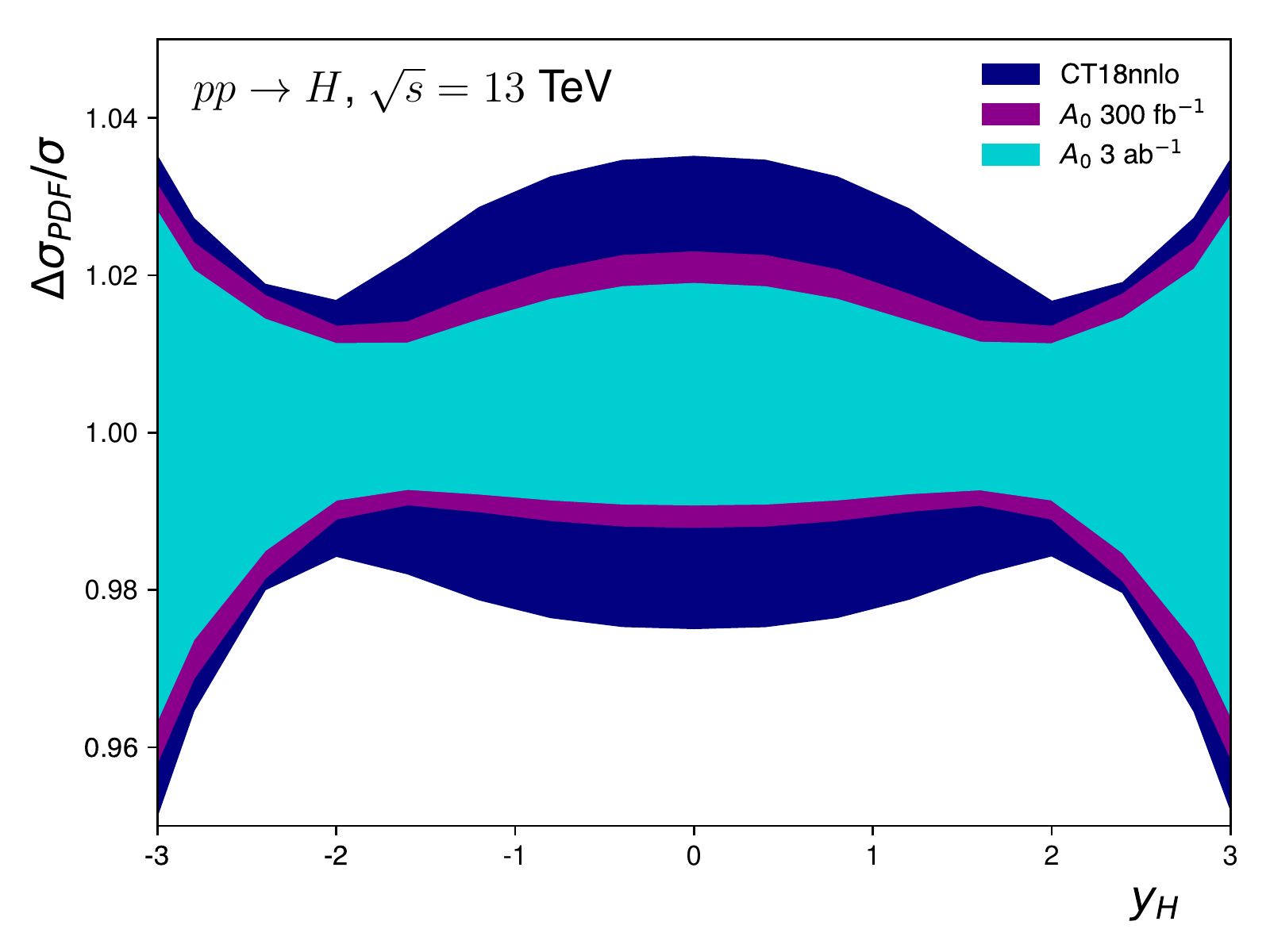}
\end{center}
\caption{Ratio of PDF uncertainties for the gluon-gluon fusion SM Higgs boson cross-section 
in $pp$ collisions at $\sqrt{s}=$13~TeV as a function of the Higgs rapidity. The red band 
shows the uncertainties of the CT18nnlo PDF set~\cite{Hou:2019efy}, reduced to 68\% CL coverage. The blue and 
green bands show the uncertainties of
the CT18nnlo including constraints from the $A_0$ measurement
and assuming 300~fb$^{-1}$ and 3~ab$^{-1}$, respectively.}
\label{fig:HiggsRapidity}
\end{figure}

The effect of the longitudinally polarized coefficient on the $Q^2$ = 10$^4$ GeV$^2$ gluon PDF near $ x \sim 10^{-2}$ 
will influence the Higgs boson cross section. 
To study this, we compute  SM Higgs boson production  in the gluon fusion  mode for $\sqrt{s}=13$ TeV $ pp$ collisions, using the MCFM code~\cite{Campbell:2010ff,Campbell:2019dru}  at NLO in QCD perturbation theory. We evaluate PDF uncertainties on the Higgs cross section including constraints from $A_0$ profiling. 
The results are given in Fig.~\ref{fig:HiggsRapidity} versus the Higgs boson rapidity $y_H$. 
We see that in the region $ - 2 \ltap y_H \ltap  2 $ the uncertainty is reduced by about 30 - 40~$ \% $ in the Run III scenario, and a further reduction to about 50~$ \% $ takes place in the HL-LHC scenario.

\begin{figure}[t!]
\begin{center}
\includegraphics[width=0.43\textwidth]{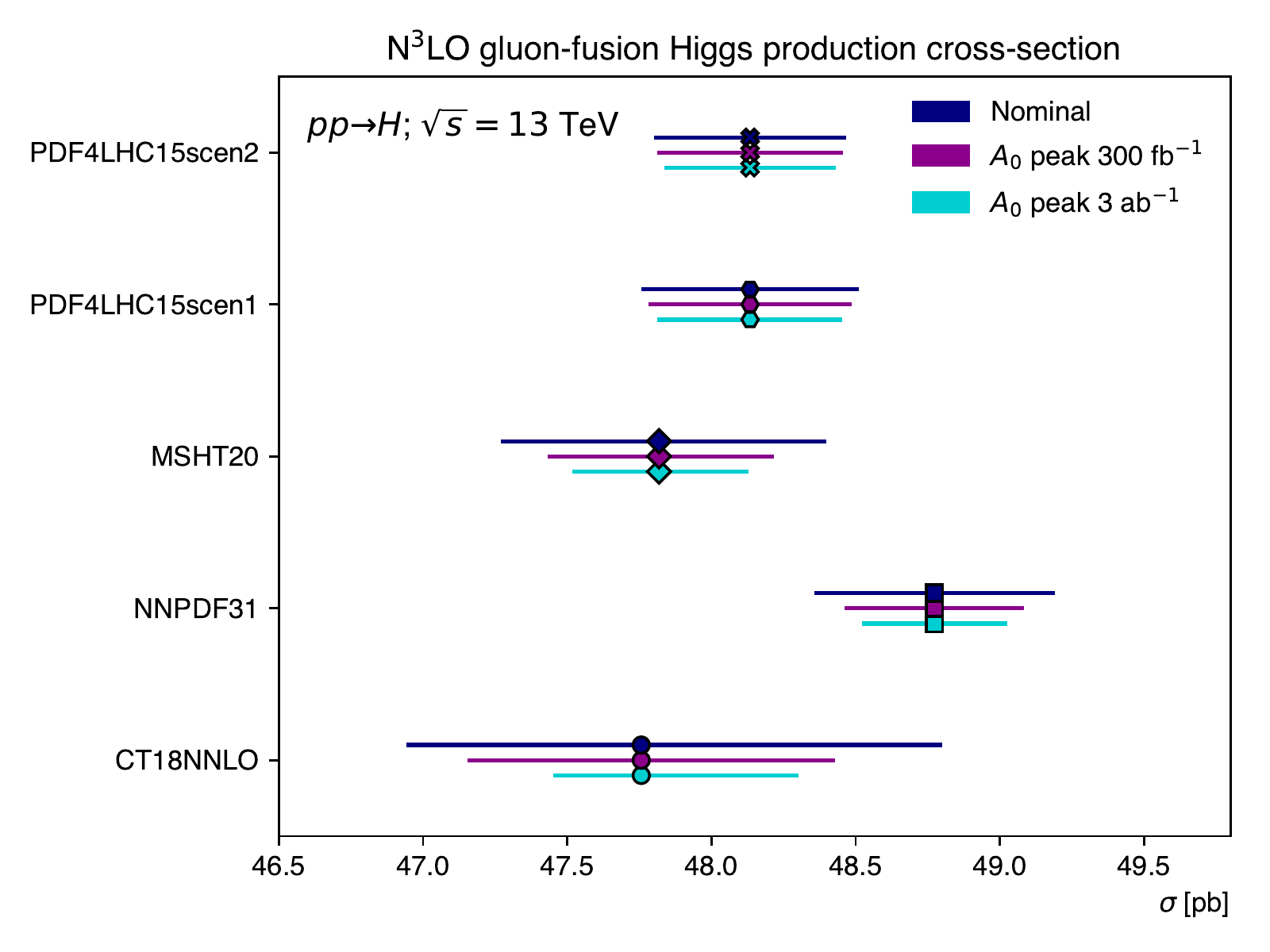}
\end{center}
\caption{The gluon-gluon fusion Higgs boson production cross-section at N$^3$LO for different PDFs, showing the 
uncertainty from PDFs and their expected reduction including constraints from the $A_0$ measurement 
assuming 300~fb$^{-1}$ and 3~ab$^{-1}$, respectively.}
\label{fig:HiggsSigma}
\end{figure}

We next perform a higher-order N$^3$LO calculation for the  Higgs boson total cross section 
using the code  {\tt{ggHiggs}}~\cite{Bonvini:2014jma,Bonvini:2018ixe}. In Fig.~\ref{fig:HiggsSigma}, we report the result for the cross section and its uncertainty in the cases of the current CT18nnlo~\cite{Hou:2019efy},  NNPDF3.1nnlo~\cite{Ball:2017nwa} and MSHT20nnlo~\cite{Bailey:2020ooq} global sets as well as projected sets, based on complete LHC data sample~\cite{Khalek:2018mdn}. 
The  PDF4LHC15scen1/2 sets, which are PDF projections including HL-LHC pseudodata, also show a smaller, but not negligible, reduction in uncertainties. 
Notwithstanding the numerical differences, the behavior is qualitatively similar for the different sets, and  provides further support at N$^3$LO to the picture given in Fig.~\ref{fig:HiggsRapidity} for the NLO Higgs boson rapidity cross section. 

The results above for the Higgs boson production cross section have been obtained using the DY angular coefficient for longitudinal electroweak boson polarization in the mass region near the $Z$-boson mass (top panel in Fig.~\ref{fig:A0_13TeV_contributions}). 
We stress that the same approach,  extended to mass regions away from the $Z$ peak, has the potential to provide complementary physics information.  For instance, high-mass DY angular distributions allow the region of larger $x$ momentum fractions to be accessed and will be relevant for associated Higgs boson production with a gauge/Higgs boson or heavy-flavour quarks. Conversely, we have noted earlier 
that  measurements of $A_0$ at low masses (center panel in Fig.~\ref{fig:A0_13TeV_contributions})  may be used to probe $p_T$ dependent gluon PDF effects, and this will impact the Higgs boson $p_T$ spectrum for low transverse momenta. The extension to low masses can further provide a handle on the 
small-$x$ regime~\cite{Bonvini:2018ixe,Hautmann:2002tu} of Higgs boson production relevant to the highest energy frontier. 

We have so far exploited the sensitivity of the longitudinal electroweak boson polarization to the gluon PDF and singled out $A_0$ as a perturbatively stable observable, which can be built via diagonal elements of the polarization density matrix and is measurable via the lepton polar angle $\theta$. 
This can be generalized as further sensitivity may arise from off-diagonal density matrix elements via longitudinal-transverse interferences, such as the parity-conserving $A_1$ and parity-violating $A_3$ coefficients~\cite{Aad:2016izn},  
which can be accessed by measuring also the lepton azimuthal angle $\phi$. These coefficients are generally smaller than $A_0$ and with a milder $p_T$ dependence, but provide a more pronounced $Y$ rapidity dependence. Moreover, starting at order $\alpha_s^2$ one may investigate additional handles from violation of the Lam-Tung relation~\cite{Lam:1978pu}, $A_2 \neq A_0$, and from the $T$-odd coefficients $A_5$, $A_6$, $A_7$~\cite{Aad:2016izn}.

{\it Conclusion.} We have proposed the systematic use of electroweak gauge boson polarization in charged lepton-pair 
hadro-production to investigate gluon-initiated  processes and the associated non-perturbative QCD contributions.

We have illustrated this by studying 
the implications of precise measurements of  
the  angular coefficient  
$A_0$  
near the $Z$-boson mass scale 
on the theoretical predictions for the Higgs boson production cross section, exploiting the coupling of the longitudinal 
polarization to the gluon PDF through radiative contributions in  $\alpha_s$. 

Our results open 
a new area of phenomenological studies on connections between the gauge  
and Higgs sectors of the SM, as further aspects may be investigated  
via generalization 
to the full structure 
of lepton angular distributions,   including polarization interferences,  and to mass regions far away from the  $Z$-boson peak.

{\it Acknowledgments}. We thank H.~Abdolmaleki,
A.~Armbruster, V.~Bertone, S.~Camarda, A.~Cooper-Sarkar, D.~Britzger, L.~Harland-Lang, A.~Huss, F.~Olness and other 
 {\tt{xFitter}} developers for useful conversations and advice. 
The work of JF has been supported by the BMBF under contract 05H15PMCCA and the DFG 
through the Research Training Network 2149 ``Strong and weak interactions - from hadrons to dark matter" and by STFC under the consolidated grant ST/T000988/1.

\section{Supplementary material} 

\subsection{Fit to ATLAS 8 TeV data} 

To validate our  theoretical framework based on the implementation of the NLO  {\tt{MadGraph5\_aMC@NLO}} $A_0$ calculation into the  fit platform {\tt{xFitter}}, 
we consider the $\sqrt{s} = 8$ TeV ATLAS measurements~\cite{Aad:2016izn}  for the angular coefficient $A_{0}$, using the
unregularised data in the three rapidity regions 
and removing the bins with $p_T<11.4$~GeV. 

We perform NLO fits which include the covariance matrices of the experimental and PDF uncertainties, for the cases CT18nnlo~\cite{Hou:2019efy}, 
NNPDF3.1nnlo~\cite{Ball:2017nwa}, ABMP16nnlo~\cite{Alekhin:2017kpj},  HERAPDF2.0nnlo~\cite{Abramowicz:2015mha} and MSHT20nnlo~\cite{Bailey:2020ooq}. The results for the $\chi^2$  values are reported in Tab.~\ref{chi2_8TeV} for each of the PDF sets, showing a very good description of  data for all sets.
 
\begin{table}[!th]
\centering
\begin{tabular}{|c|c|}
\hline
PDF set & Total $\chi^2$/d.o.f.\\
\hline
CT18NNLO & 59/53 \\
\hline
CT18Annlo & 44/53 \\
\hline
NNPDF31{\_}nnlo{\_}as{\_}0118{\_}hessian & 60/53 \\
\hline
ABMP16{\_}5{\_}nnlo & 62/53 \\
\hline
MSHT20nnlo{\_}as118 & 59/53 \\
\hline
HERAPDF20{\_}NNLO{\_}EIG & 60/53\\
\hline
\end{tabular}
\caption{The $\chi^2$ values per degrees of freedom from NLO fits to $A_{0}$ data~\cite{Aad:2016izn} 
using  {\tt{xFitter}}, for different collinear PDF sets. PDF uncertainties are evaluated at the 68\% CL.}
\label{chi2_8TeV}
\end{table} 

\subsection{Profiling in the low-mass region and forward region} 

\begin{figure}[t!]
\begin{center}
\includegraphics[width=0.23\textwidth]{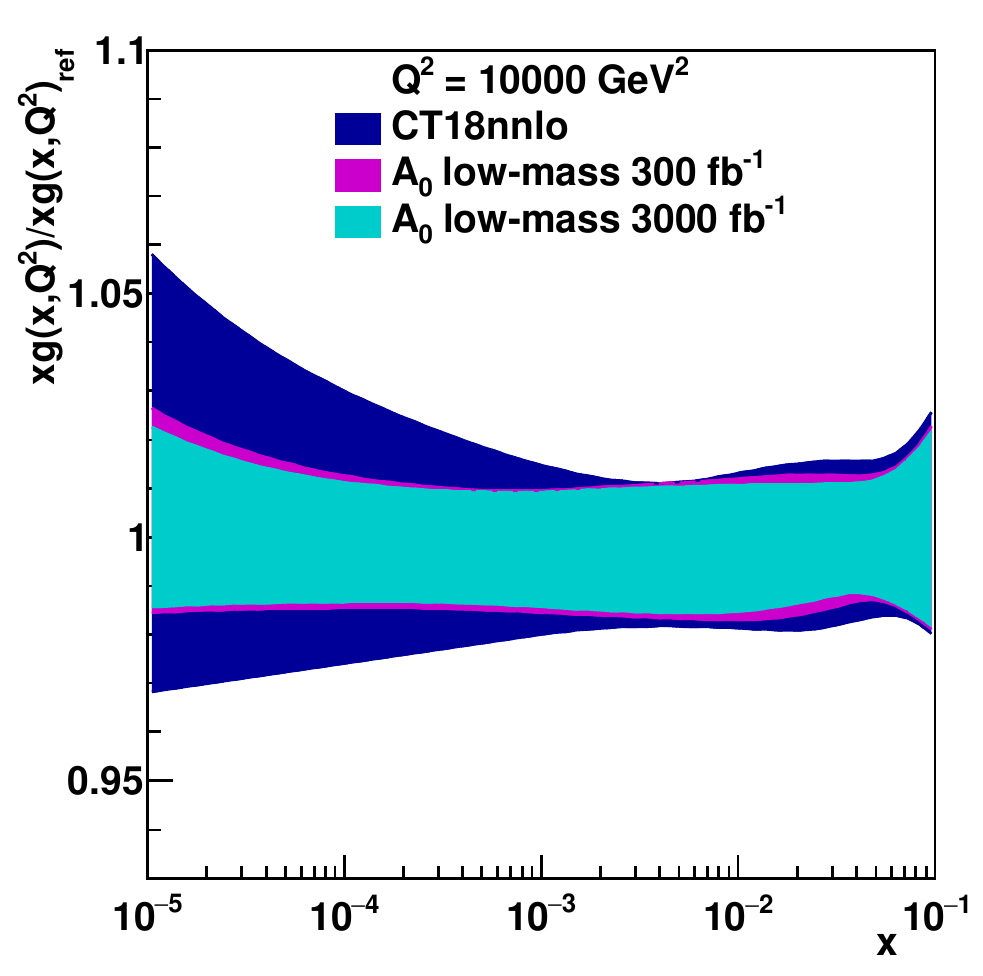}
\includegraphics[width=0.23\textwidth]{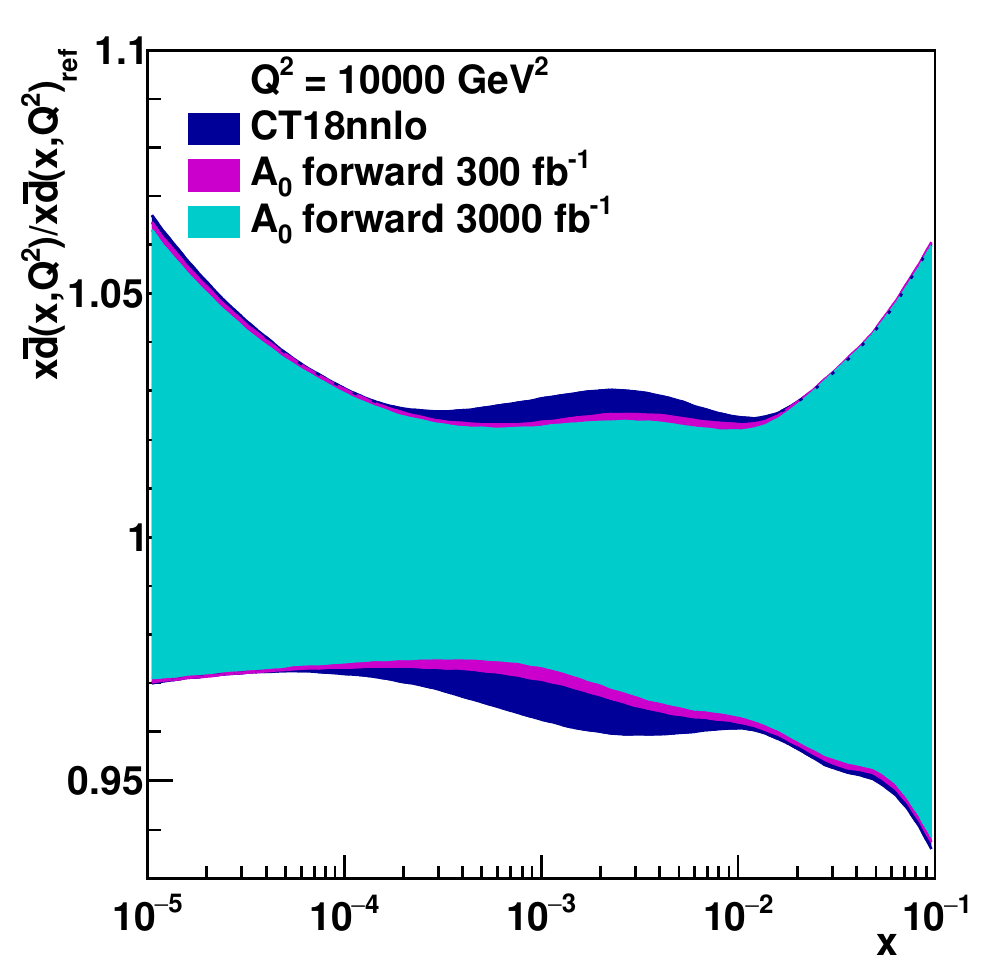}
\caption{Original CT18nnlo~\cite{Hou:2019efy} (red) and profiled distributions using $A_0$ pseudodata corresponding to integrated luminosities of 300~fb$^{-1}$ (blue) and 3~ab$^{-1}$ (green) for the low mass (left) and LHCb phase space (right). 
  Results for gluon ($xg$) and $d$-type ($x\bar{d}$) sea-quark densities are shown for $Q^{2}$ = 10$^{4}$~GeV$^{2}$. 
Bands represent PDF uncertainties, shown at the 68\% CL.}
\label{fig:CT18_13TeV_Second_appendix}
\end{center}
\end{figure}

Analogously to the profiling analysis of Fig.~\ref{fig:CT18_13TeV_Second} in the $Z$-boson peak region, 
we perform the profiling analysis in the low mass region  4~GeV $< M <$ 8~GeV 
and forward rapidity region 2.0 $< |Y| <$ 4.5. We report the results in Fig.~\ref{fig:CT18_13TeV_Second_appendix}. 

Given that triggering in this kinematic domain is challenging we consider 
that only a 1\% fraction of the data will be recorded,
thereby reducing the effective
luminosity by this factor.
The largest reduction of uncertainties observed in this case
for the gluon distribution is at $x<0.001$. For the LHCb phase space, the most relevant improvements
are for the sea quark PDFs, e.g. $\bar{d}$ at $x\sim 0.001$.

\subsection{Gluon-gluon luminosity}

We compute the gluon-gluon luminosity as a function of invariant mass to assess the reduction of uncertainties as a result of PDF profiling based on the 
longitudinally polarized angular coefficient $A_0$.

\begin{figure}
\begin{center}
\includegraphics[width=0.43\textwidth]{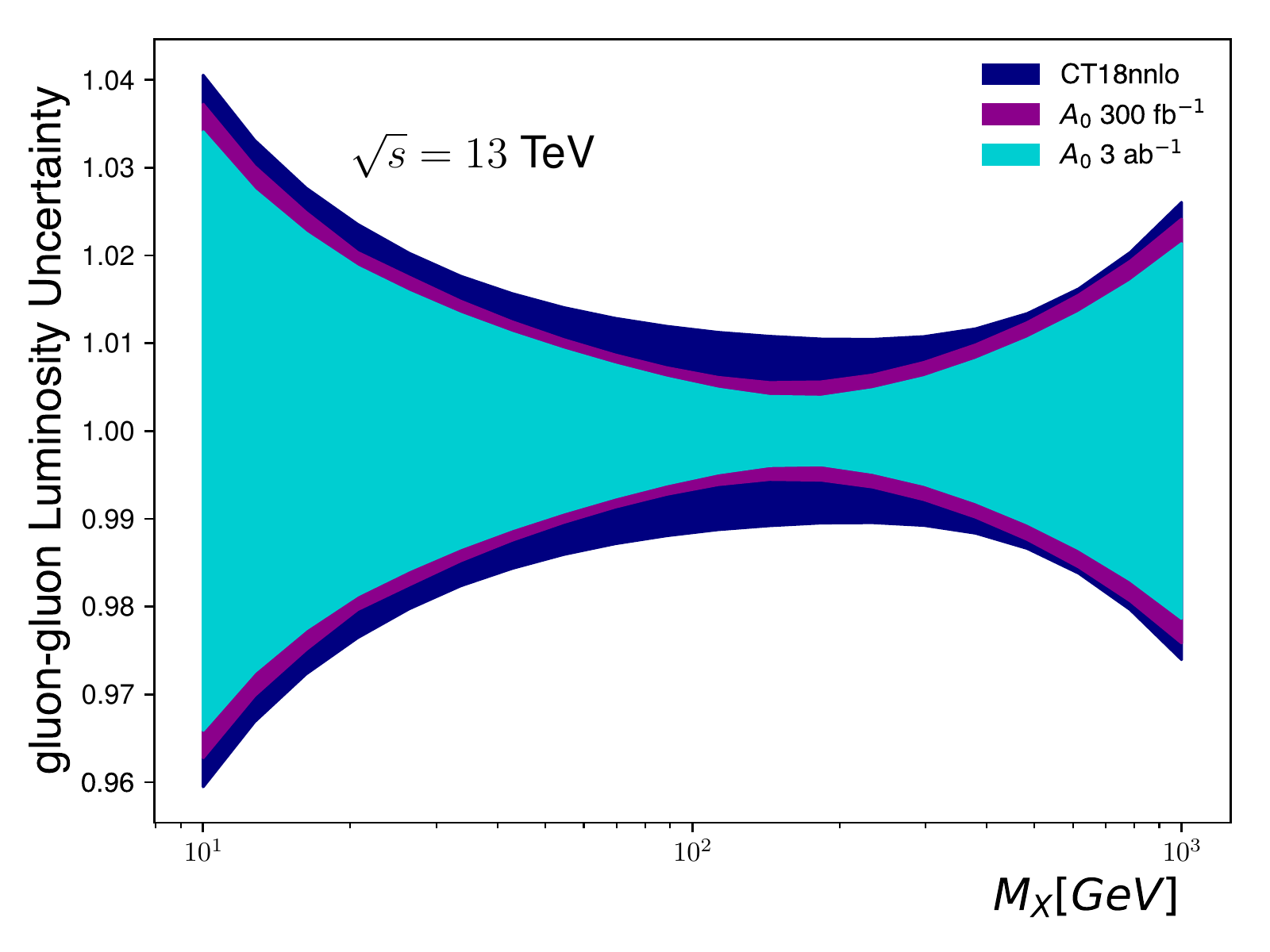}
\end{center}
\caption{Ratio of PDF uncertainties for the gluon-gluon luminosity evaluated at $\sqrt{s}=$13~TeV. The red band  shows the uncertainties of the CT18nnlo PDF set~\cite{Hou:2019efy}, reduced to 68\% CL coverage. The blue and 
green bands show the uncertainties of the CT18nnlo including constraints from the $A_0$ measurement and assuming 300~fb$^{-1}$ and 3~ab$^{-1}$, respectively.}
\label{fig:ggLuminosity}
\end{figure}

In Fig.~\ref{fig:ggLuminosity} we show the PDF uncertainties in the gluon-gluon luminosity evaluated at $\sqrt{s}=$13~TeV and computed with CT18nnlo, as well as including constraints from $A_0$ profiling. PDF uncertainties are halved in the range 100 $< M_{X} <$ 200~GeV in the HL-LHC scenario.

\subsection{Profiling projected PDFs based on complete LHC data sample}

\begin{figure}
\begin{center}
\includegraphics[width=0.45\textwidth]{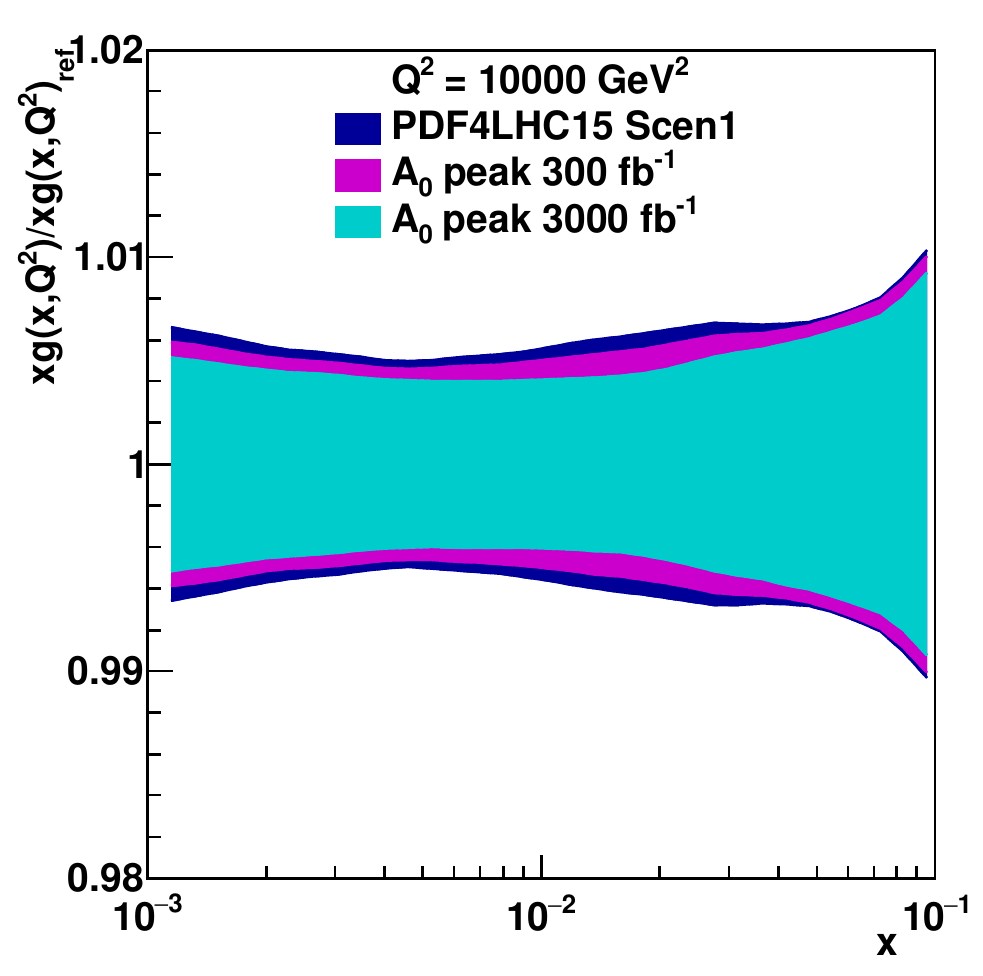}
\end{center}
\caption{Original PDF4LHC15 Scenario 1~\cite{Khalek:2018mdn} (red) and profiled distributions using $A_0$ pseudodata corresponding to integrated luminosities of 300~fb$^{-1}$ (blue) and 3~ab$^{-1}$ (green) for   80~GeV $< M <$ 100~GeV and $|Y|<3.5$. 
  Result for gluon ($xg$) is shown for $Q^{2}$ = 10$^{4}$~GeV$^{2}$. 
Bands represent PDF uncertainties, shown at the 68\% CL.}
\label{fig:PDF4LHC15_appendix}
\end{figure}

Paralleling the profiling analysis of Fig.~\ref{fig:CT18_13TeV_Second}, in Fig.~\ref{fig:PDF4LHC15_appendix} we report the profiling analysis which uses projected PDFs based on complete LHC data sample as input~\cite{Khalek:2018mdn}. Remarkably, 
a reduction in the gluon PDF uncertainty is still visible. This picture is consistent with what already depicted in 
Fig.~\ref{fig:HiggsSigma}. We further observe that,  
in the PDF4LHC15Scen1 profiling scenario of Fig.~\ref{fig:PDF4LHC15_appendix},   the   300 fb$^{-1}$ to 3 ab$^{-1}$ gain relative to the 
current-to-300 fb$^{-1}$ gain is more significant  than in the case of the CT18nnlo~\cite{Hou:2019efy} 
profiling scenario in the top left panel of Fig.~\ref{fig:CT18_13TeV_Second}.

\end{document}